\documentclass[a4paper,11pt]{article}

\pdfoutput=1 

\usepackage{jcappub} 

\usepackage{amsmath}
\usepackage{amssymb}
\usepackage{soul}
\usepackage{ulem}

\newcommand{\rmd}{{\rm d}} 

\title{Can varying the gravitational constant alleviate the tensions ?}

\author[a]{Z. Sakr}
\author[b]{and D. Sapone}

\affiliation[a]{Physics Department, University of Saint Joseph, Beirut, Lebanon}
\affiliation[b]{Departamento de F\'isica, FCFM, Universidad de Chile, Blanco Encalada 2008, Santiago, Chile}

\emailAdd{ziad.sakr@net.usj.edu.lb}
\emailAdd{dsapone@ing.uchile.cl}

\abstract{
Constraints on the cosmological concordance model parameters from observables at different redshifts are usually obtained using the locally measured value of the gravitational constant $G_N$. Here we relax this assumption, by considering $G$ as a free parameter, either constant over the redshift range or dynamical but limited to differ from fiducial value only above a certain redshift. Using CMB data and distance measurements from galaxy clustering BAO feature, we constrain the cosmological parameters, along with $G$, through a MCMC bayesian inference method. Furthermore, we investigate whether the tensions on the matter fluctuation $\sigma_8$ and Hubble $H_0$ parameter could be alleviated by this new variable. We used different parameterisations spanning from a constant $G$ to a dynamical $G$. In all the cases investigated in this work we found no mechanism that alleviates the tensions when both CMB and BAO data are used with $\xi_{\mathrm{g}} = G / G_N$ constrained to 1.0$\pm0.04$ (resp. $\pm0.01$) in the constant (resp. dynamical) case. Finally, we studied the cosmological consequences of allowing a running of the spectral index, since the later is sensitive to a change in $G$. For the two parameterisations adopted, we found no significant changes to the previous conclusions.

}

\begin{document}
\maketitle
\flushbottom

\section{Introduction}

The $\Lambda$CDM concordance model has been successful in accommodating with most of nowadays cosmological probes like the cosmic microwave background (CMB) temperature and polarisation power spectrum, weak lensing correlations, galaxy clustering and supernova's Hubble diagram. Recent measurements of the aforementioned observables, Planck \cite{Planck:2018vyg}, KIDS \cite{KiDS:2020suj}, eBOSS \cite{deMattia:2020fkb}, DES \cite{Abbott:2017wau} and PANTHEON survey \cite{SNLS:2011lii} have succeeded in putting strong constraints on the parameters of the $\Lambda$CDM model with near percent precision. However, these constraints have been obtained with some prior assumptions, among them, a fixed value for the Newtonian Gravitational coupling ($G_N$) treated as a constant both in the Newton’s or the general relativity theory which the Committee on Data for Science and Technology (CODATA) \cite{Mohr:2015ccw} recommends value as $G_N = 6.67430 \times 10^{-11}\,\rm{m}^3\,\rm{kg}^{-1}\,\rm{s}^{-2}$. This is strongly justified from the stringent constraints on $G$ possible variation in time or space, obtained from experiments in lab and solar system, see
\cite{2015EL....11010002A,Schlamminger:2015hqa, Li2018MeasurementsOT,Dai:2021jnl}. Further astrophysical studies  corroborated this assumption after attempts to test deviation from fiducial of a time-varying $G$ using Type Ia supernovae \cite{Mould:2014iga, Sapone:2020wwz} or Tully-Fisher relation data \cite{Alestas:2021nmi}, found consistency with the gravitational constancy hypothesis.

However, unlike the lab or solar system experiments, those on larger astronomical scales, do not operate on $G$ itself and the analysis rather include proxies to mass of interacting bodies, leaving room for possible degeneracies through which a variation from fiducial $G_N$ remains possible. This is even more the case when we explore cosmological scales at higher redshifts, far from our local universe. That is why some have attempted to revisit the different cosmological studies while relaxing the $G$ value in order to test the impact of such change on the cosmological parameters constraints and to infer $G$ independently from local measurements. The latter goal has some theoretical justification, since many modified gravity theories predict a time-dependent $G$ as an attempt to naturally model the accelerated expansion of the universe, \cite{1935rgws.book.....M, Dirac:1937ti,1961PhRv..124..925B,Wetterich:2013jsa, Burrage:2020jkj,Ballardini:2020iws,Braglia2020LargerVF,Will:2014kxa, 2021EPJP..136..143F, Kaluza:1921tu,Klein:1926fj, ArkaniHamed:1998rs, Randall:1999ee,Randall:1999vf, Dvali:2000hr, Damour:1990tw, Capistrano:2020awr, Bronnikov:2020zre, Hanimeli:2019wrt,Nesseris:2017vor,Ballardini:2021evv,Abadi:2020hbr}.

Another motivation, related to the aforementioned impact on the cosmological parameters constraints, is justified by the growing evidence of two discrepancies found lately on the matter fluctuation parameter $\sigma_8$ and the Hubble parameter $H_0$, between their values constrained from deep universe probes like the CMB correlations in comparison with those from local ones, like cluster counts or weak lensing correlations for the $\sigma_8$ inference \cite{Hoekstra:2015gda,Planck:2013shx} or the Cepheids luminosity distance for $H_0$ \cite{Riess:2019cxk}. There has been attempts to alleviate the tension for, either $\sigma_8$ \cite{Sakr:2018new,Ilic:2019pwq}, or the Hubble parameter, for each alone or for both at the same time (see \cite{DiValentino:2021izs} and references therein), with relative successes. Especially when combining with the baryonic acoustic oscillation (BAO) angular distance probe, where it was shown that almost all the gain in reducing the discrepancy between deep and local probes vanishes when the BAO constraints were added see e.g. \cite{Jedamzik:2020zmd}. This was also found by similar attempts to fix the discrepancy with some of the models we mentioned earlier \cite{Ballardini:2020iws,Braglia2020LargerVF,Capistrano:2020uac,2020arXiv200604273S,Gupta:2021tma}. Other approaches that were explored meant to have a transition of $G$ at a particular redshift, see \cite{Kazantzidis:2020tko, Marra:2021fvf, Alestas:2021luu} or one in $H_0$ in general \cite{Dainotti:2021pqg}. 

In this work we use cosmological observables like CMB temperature and polarisation to constrain deviations of $G$ from fiducial but we do not consider modification to $G$ as a result of a particular modified gravity. We rather choose to test deviation and extension to the fiducial fixed G N value with general phenomenological parameterisations and try to account for the consequences of such parameterisation on the observables even on the microphysics involved.

Similar studies on the deviation from fiducial $G$ in cosmology have been considered, for instance: in \cite{Zahn:2002rr}, it was provided a method to constrain the gravitational constant, showing that a degeneracy between its effect on the expansion rate and the primordial spectrum can be broken by measuring CMB polarisation along with its temperature correlations; in \cite{Umezu:2005ee} were presented cosmological constraints on deviations of Newton’s constant at large scales, analysing the cosmic microwave background (CMB) anisotropies and primordial abundances of light elements synthesised by big bang nucleosynthesis; and in \cite{Galli:2009pr}, where CMB data were used together with BBN priors to constrain $G$ in models of free constant or varying G; whereas in \cite{Mould:2014iga}, SN data were used to set a limit on $G$ variation with time. These analysis were repeated later, with updated measurement of $G$ after the first Planck release \cite{Bai:2015vca}, confirming that $G/G_N$ is in agreement with being equal to one. Recently, there have been works using BNN data \cite{Alvey:2019ctk} or adding to the CMB, local data such as BAO or SNIa measurements, \cite{Wang:2020bjk}.

Here we follow the same idea by considering $G$, within general relativity theory, as a free parameter allowed to vary from the fiducial $G_N$, paying attention to its impact on all the main physical processes involved in the cosmological observables. This has a double objectives of, first exploring the consequences on the G constraints, and second its impact on the aforementioned discrepancies and the other cosmological parameters. Since it has been shown that models with late modification do not succeed in fixing the discrepancies, especially when combining with BAO constrains, we shall, as describe later, consider parameterisations where $G$ deviates from $G_N$ only at early times before reverting back to its fiducial value after a certain redshift, with the latter threshold chosen at different times outside the BAO domain presently measured at redshfits below $z \sim 2.5$.

The outline of the paper is the following. In Sect.~\ref{sect:model_prob} we discuss models for modified $G$ and implications on the cosmological probes. In Sect.~\ref{sect:data_method} we describe the datasets used and the way modifications to observables were implemented. In
Sect.~\ref{sect:G_mod_results} we present and discuss the results of a Bayesian analysis constraints and their implications on the cosmological observables, and we conclude in Sect.~\ref{sect:G_mod_concl}. 

\section{Variable G : Models and cosmological implications }\label{sect:model_prob}

We introduce a dimensionless parameter $\xi_{\mathrm{g}}$ to
quantify the potential deviation of Newton’s gravitational $G$ from the laboratory-based measurement $G_N$. We consider three phenomenological parameterisation: 
\begin{itemize}
\item constant case scaling with $\xi_{\mathrm{g}}$ 
\begin{equation}
 G = \xi_{\mathrm{g}} \times G_N\,;
 \end{equation}
\item a step like function where the parameter $\xi_{\mathrm{g}}$ is different from fiducial $G_N$, above a certain redshift $z_{tr}$,
\begin{eqnarray}G=
\begin{cases}
\xi_{\mathrm{g}} \times G_N,&~\text{for}~z>z_{tr};\\
 G_N,&~\text{for}~z<z_{tr};\\
\end{cases}
\label{equ:ztrans}
\end{eqnarray}
\item an hyperbolic tangent function, where $\delta_z$ is the width of the transition:
 \begin{equation}
 G = \left[1- \left(1 - \xi_{\mathrm{g}}\right)\left(1-0.5\right)\times \tanh\left( \frac{z_{tr}-z}{\delta_z}\right)\right] \times G_N\,.
  \end{equation} 
  \end{itemize}
To show the impact on observables like the CMB spectrum or BAO feature, we consider, without loose of generality, the first parameterisation. The Friedmann equation then becomes (working in $c = 1$ units):
  \begin{equation}
H^2=\left(\frac{\dot{a}}{a} \right)^2= \frac{8 \pi}{3} \xi_{\mathrm{g}}^2 G_N  \rho_{\rm tot}
\end{equation}
where $\rho_{\rm tot}$ is the total energy density. 
This implies that the expansion rate $H$ satisfies:
\begin{equation}
H(a,\xi_{\mathrm{g}})=\xi_{\mathrm{g}} \times H(a)\,,
\end{equation}
so that the shape of $H$ is not changed by $\xi_{\mathrm{g}}$. This serves to see the consequence on the CMB temperature power spectrum. To show its effect, we start by writing the temperature perturbation as an integral along the line of sight over the sources:
\begin{equation}
\Delta T_l\left(k,\xi_{\mathrm{g}}\right) = \int^{1}_a  {\rm d}a \; \tilde S\left(k,a,\xi_{\mathrm{g}}\right) e^{i k\cdot \hat n r\left(a,\xi_{\mathrm{g}}\right)} \tilde g\left(a,\xi_{\mathrm{g}}\right)\,, 
\label{eq:deltat_l}
\end{equation}
where $S(k,a)$ is the anisotropy source term and $r(a)$ is the distance from the observer to a point along the line of sight at the scale factor $a$, and $g(a)$ is the visibility function. Assuming, form the moment, $\xi_{\mathrm{g}}$ constant, the distance $r(a)$ reads:
\begin{equation}
r(a,\xi_{\mathrm{g}}) =\int_{a}^1 \frac{\rm{d}a}{H(a,\xi_{\mathrm{g}}) a^2}= r(a)/\xi_{\mathrm{g}}
\end{equation}
and the $\xi_{\mathrm{g}}$ parameter is factored out in Eq.~\eqref{eq:deltat_l}, satisfying $S(k,a,\xi_{\mathrm{g}}) =  S(k/\xi_{\mathrm{g}},a)$, as also shown by \cite{Zahn:2002rr}. 

Since the CMB angular power spectrum is calculated from $\Delta T_l(k,\xi_{\mathrm{g}})$ using
\begin{equation} 
C_l(\xi_{\mathrm{g}}) = \int \frac{{\rm d} k}{k} P(k_0) |\Delta_{Tl}(k,\xi_{\mathrm{g}})|^2 
\end{equation} 
where $P(k_0)$ is the power-law primordial power spectrum, we can choose to scale the latter by means of a transformation $k'=k *\xi_{\mathrm{g}}$, this will compensate the $\xi_{\mathrm{g}}$ modifications so that the CMB power spectrum ends up being invariant by our parameterisation of $G$. 

This holds if the visibility function is not affected by $\xi_{\mathrm{g}}$. However, the physics of recombination introduces a preferred timescale. This happens through the visibility function $g(a)$ written in terms of $\kappa$ (the opacity for Thomson scattering) that is dependent of $\xi_{\mathrm{g}}$ as:
\begin{equation}
g(a,\xi_{\mathrm{g}}) =-a^2H\left(a,\xi_{\mathrm{g}}\right)\frac{{\rm d}}{{\rm d}a} \exp\left(-\kappa\left(a,\xi_{\mathrm{g}}\right)\right)
\end{equation}
with
\begin{equation}
\kappa (a,\xi_{\mathrm{g}}) = \sigma_T \int_a^{1}  \frac{n_e(a,\xi_{\mathrm{g}})}{H(a,\xi_{\mathrm{g}})a^2}~ {\rm d}a,
\label{equ:opacity}
\end{equation}
\noindent 
where $\sigma_T$ is the Thomson scattering cross section, $a$ is the scale factor, with the number of free electrons $n_e$  affected by $\xi_{\mathrm{g}}$ as: 
\begin{equation}
\frac{dx_{e}}{da} = \frac{C_{r}}{a~H(a,\xi_{\mathrm{g}})} \left[ \beta(T_{b})(1-x_{e}) - n_{\rm H} \alpha(T_{b})x_{e}^{2} \right] \,.
\end{equation}
Here, $n_{\rm H}$ is the total number density of Hydrogen nuclei, $x_e=n_e/n_{\rm H}$ is the ionisation fraction, $T_b$ is the baryon temperature, $\beta(T_{b})$ is the collisional ionisation rate from the ground state, $\alpha(T_{b})$ the recombination rate to excited states and $C_r$ is the Peebles correction factor to account the presence of non-thermal Lyman-$\alpha$ resonance photons.
We also account for modifications in the ionisation fraction of Hydrogen $x_p$ and Helium $x_{\rm HeII}$. 
Based on \cite{Seager:1999bc}, we implemented the changes of $G$ in the Boltzman code (see Sec.~\ref{sect:data_method}) by solving the following differential equations as a function of $\xi_{\mathrm{g}}$:
\begin{eqnarray}
\frac{\rmd x_p}{\rmd a}&=&\frac{f_1(x_e,x_p,n_{\rm H},T_{\rm M})}{a~H(\xi_{\mathrm{g}},a)},\\
\frac{\rmd x_{\rm HeII}}{\rmd a}&=&\frac{f_2(x_e,x_{\rm HeII},n_{\rm H},T_{\rm M})}{a~H(\xi_{\mathrm{g}},a)},\\
\frac{\rmd T_{\rm M}}{\rmd a}&=&\frac{f_3(x_e,T_{\rm M},T_{\rm R})}{a~H(\xi_{\mathrm{g}},a)}+2aT_{\rm M},
\end{eqnarray}
where $T_{\rm M}$ ($T_{\rm R}$) is the matter (radiation) temperature and where the specific expressions of $f_1$, $f_2$ and $f_3$, independent of $\xi_{\mathrm{g}}$ are given in \cite{Seager:1999bc}. 

There is still an effect from the change in the Poisson equation that relates the gravitational potential $\Phi$ to the density contrast $\delta$ 
\begin{equation}
\nabla^2\Phi = - 4 \pi\xi_{\mathrm{g}} G_N \rho\,\delta\,,
\end{equation}
which influences the CMB anisotropy on large scale through the Integrated Sachs Wolfe effect (ISW). The linear ISW temperature shift along direction $\hat n$ is calculated from the time-dependent gravitational potential $\Phi$ 
\begin{equation}
\label{eq:ISW_definition2}
\frac{\Delta T_{\rm ISW}}{\overline{T}}({\hat n}) = -2\int_a^{1} a\left(1-f(a)\right)\Phi\left({\hat n},a\right)\,{\rm d} a\,,
\end{equation}
where $f= \mathrm{d}\ln D /\mathrm{d}\ln a$ is the linear growth rate of structure and $D(a) = \delta(a)/\delta(1)$ is the growth of density contrast $\delta(a)$. The equation of motion for $\delta(a)$ is:
\begin{equation}\label{eqn:ldeq}
 \delta^{\prime\prime}+\left(\frac{3}{a}+\frac{E^\prime}{E}\right)\delta^\prime-\frac{3}{2}\frac{\Omega_{\mathrm{m},0}}{a^5E^2}\delta=0\
\end{equation}
where prime denotes derivation with respect to a and $E^2(a)=H^2(a,\xi_{\mathrm{g}})/H^2_0 $. Thus, $H_0$, $\Omega_{\mathrm{m},0}$ and, $\xi_{\mathrm{g}}$ will be constrained from CMB through this effect. 

The variation of $G$ will also affect the polarisation of the CMB, allowing us to use the polarisation spectrum and cross correlations with the temperature fluctuations spectrum of the CMB to constrain $\xi_{\mathrm{g}}$. 

To understand the effect of $\xi_{\mathrm{g}}$ on the polarisation, let us consider that the degree of linear polarisation, $\Delta P$, is defined in terms of the Stokes parameters $Q$ and $U$ of the CMB radiation. Choosing two orthogonal directions basis, into which the intensity is projected, so that  $U = 0$, we can limit the study to the amplitude of the $Q$ Stokes parameter produced by a single Fourier mode $k$ \citep{Zaldarriaga:1995gi} yielding:
\begin{equation}
Q \propto c_s\,k\,\delta\tau_D \sin (k c_s \tau_D) g(a,\xi_{\mathrm{g}})
\label{eq:q-stokes}
\end{equation}
where $g(a,\xi_{\mathrm{g}})$ is the visibility function, $c_s$ is the photon-baryon sound speed, $\tau_D$ is the conformal time ( $\tau = \int dt~a_0 /a$) when $G$ peaks and $\delta \tau_D$ its width. The extra $\delta \tau_D$ in Eq.~\eqref{eq:q-stokes} implies that, if the visibility function is wider, the photons will travel on average longer between their last scatterings, enhancing the quadrupole anisotropy, thus increasing the polarisation. This effect will induce a characteristic scale $k^*$, above which the effects on the polarization is larger.

A change in the value of $G$ could also impact the reionisation process influencing the CMB observations. During the reionisation epoch, CMB photons are Thomson scattered by the free electrons produced, which suppress the amplitude of the observed primordial anisotropies. Additionally, large-scale polarisation of the CMB radiation could be produced above the earlier ones during recombination. However the reionisation process is complicated and includes many astrophysical variables whose study is outside the scope of this work. Therefore, we consider only model with simple almost instant parameterisazation so that any influence of $G$ on time and volume will not manifest, as well as any gravitational effect degenerate with $G$ will not influence the physics of reionisation from forming stars. In practice the ionisation fraction $x_e$ is parameterised as a $\tanh$ function with a width $\Delta \, z=0.5$ around the reionisation redshift \cite{Lewis:2008wr}.

\begin{figure}[!t]
\centering
\includegraphics[scale=0.35]{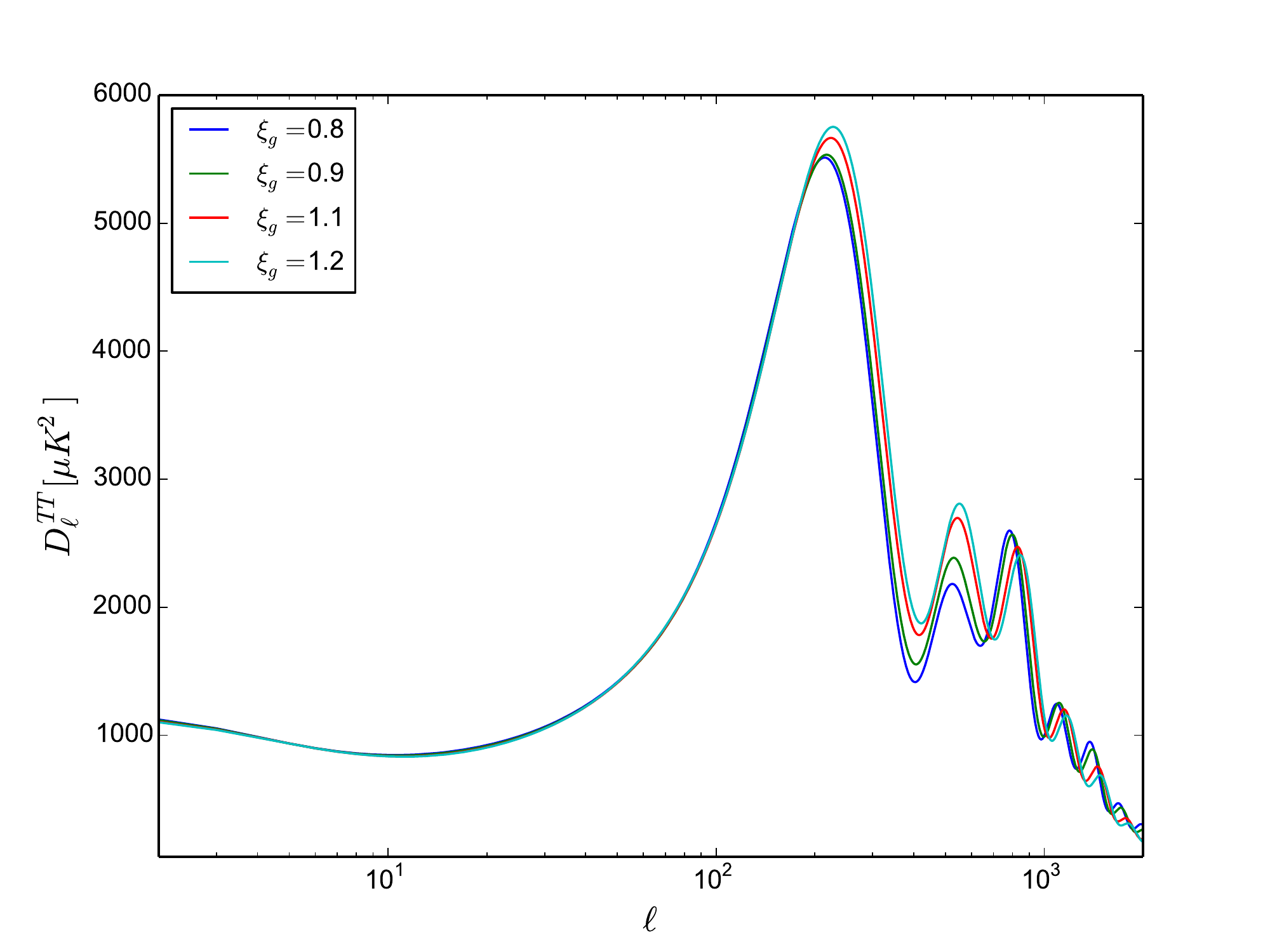}
\includegraphics[scale=0.35]{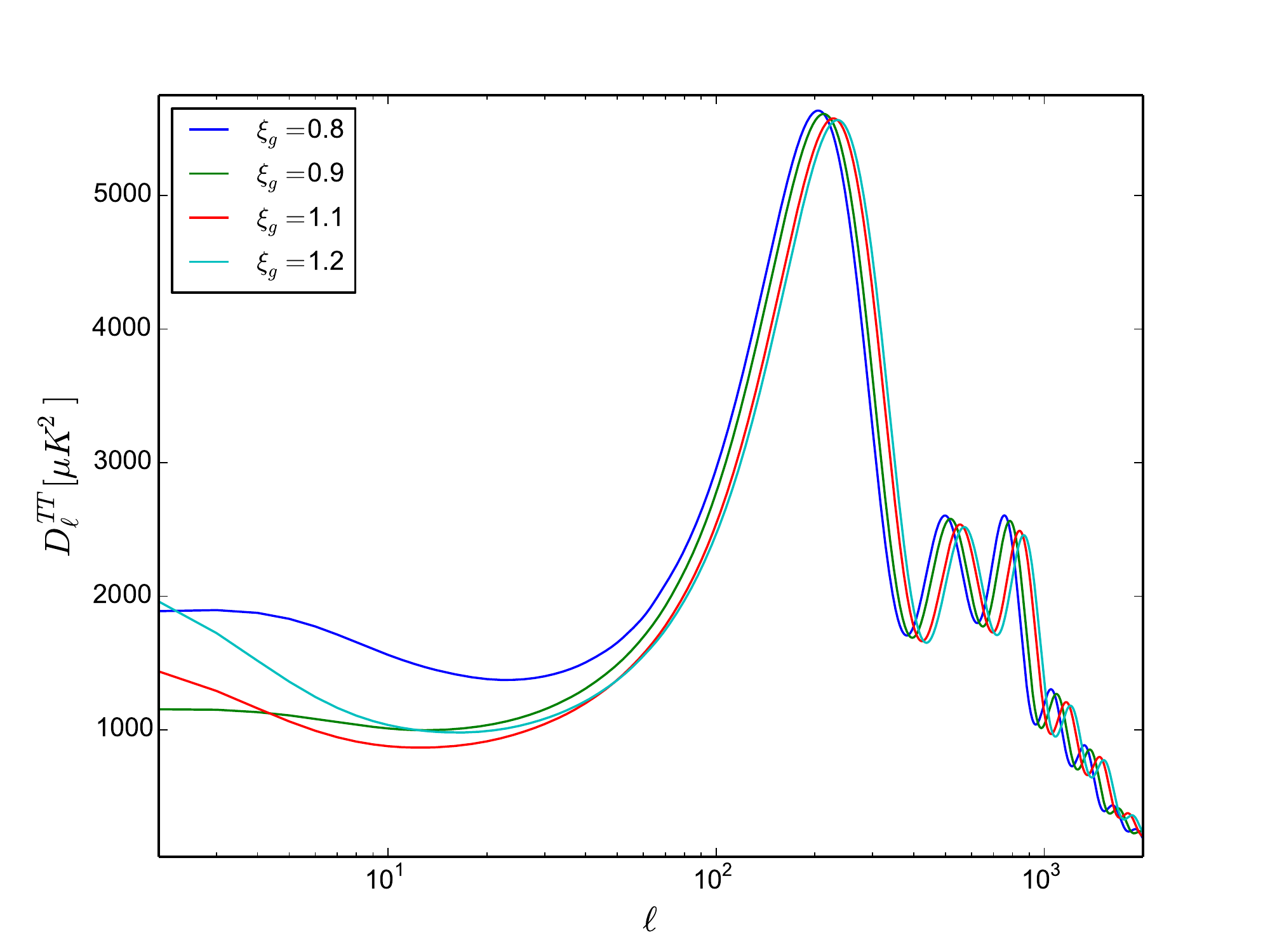}\\
\includegraphics[scale=0.35]{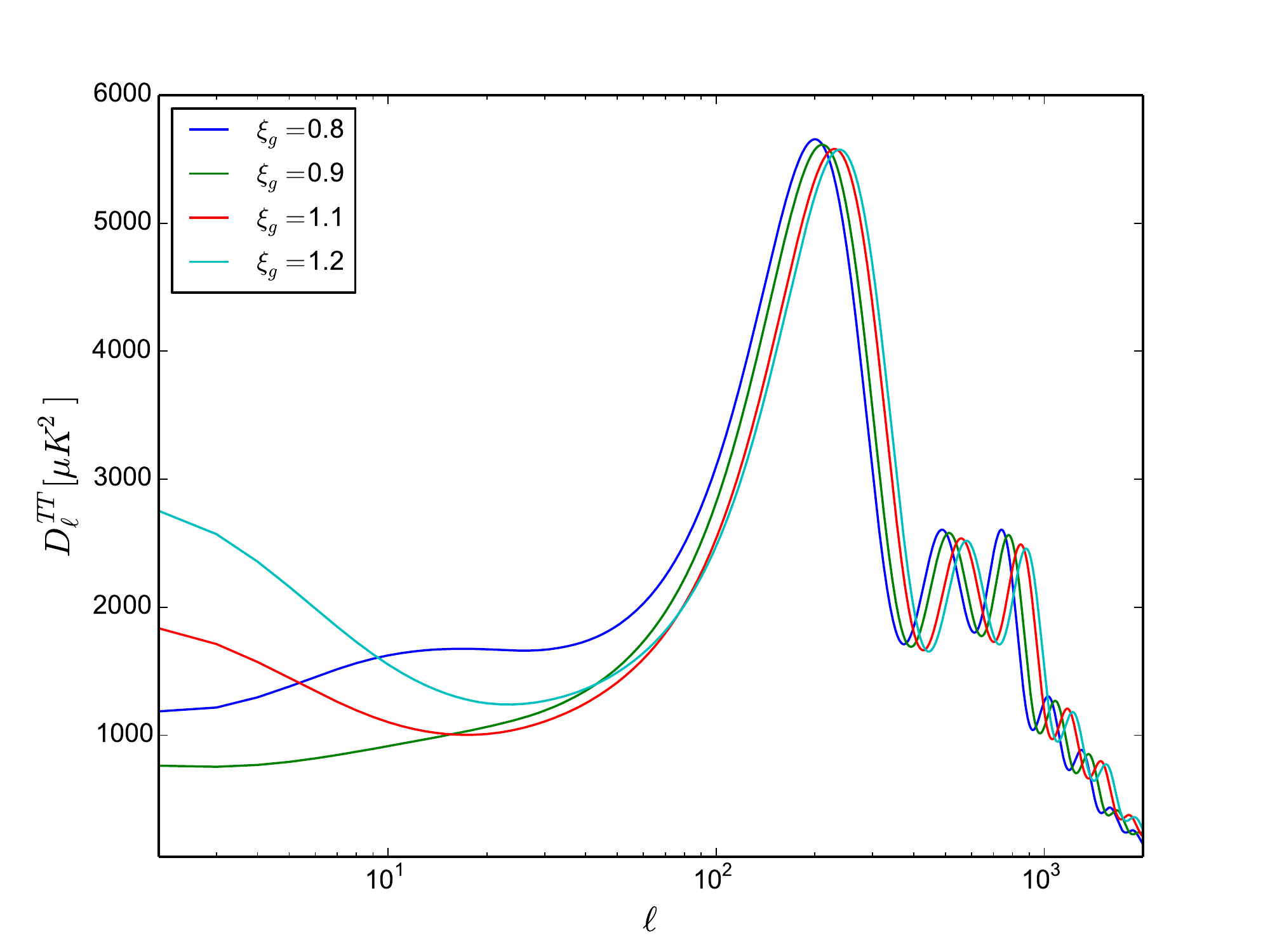}
\includegraphics[scale=0.35]{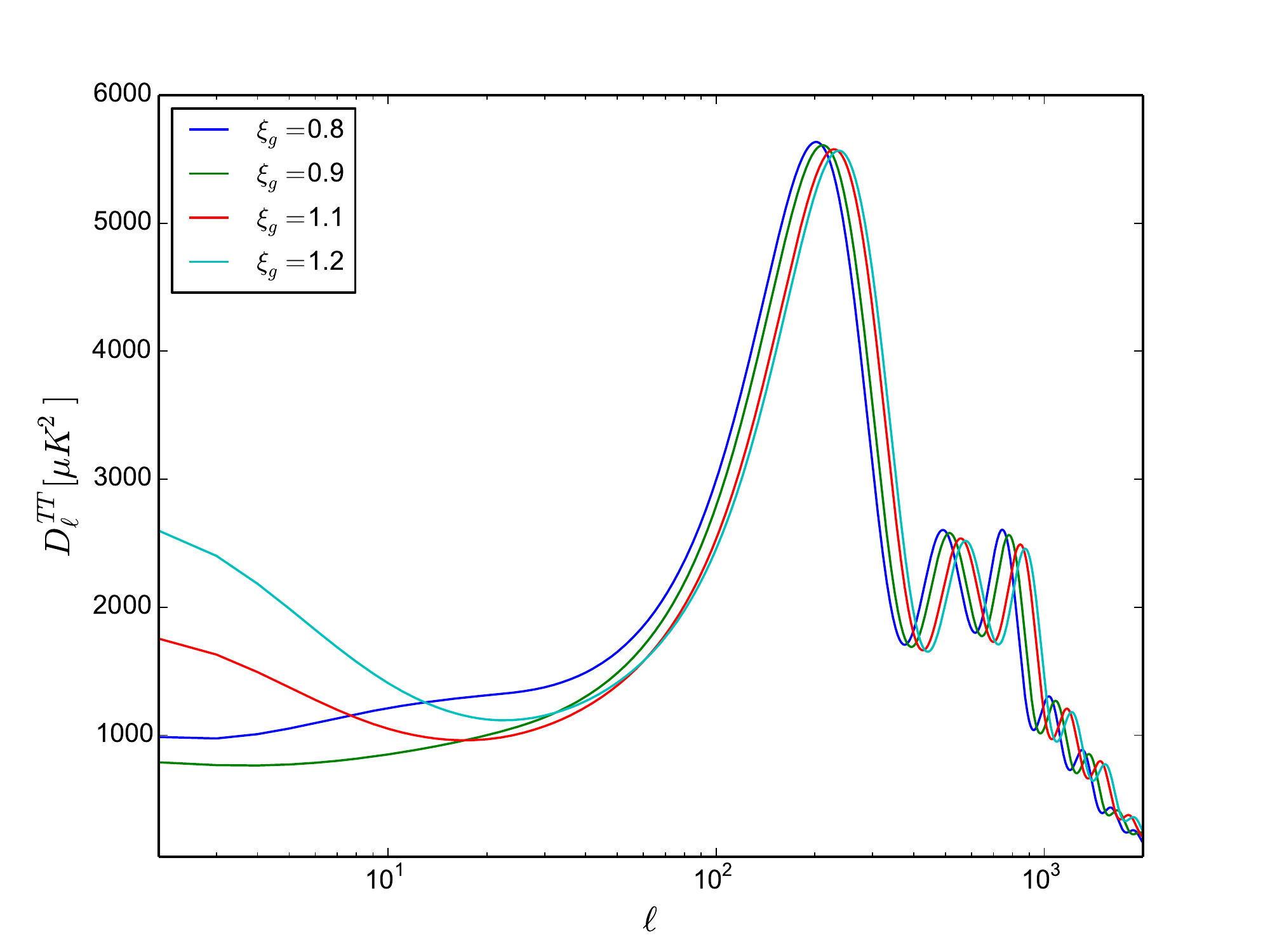}
\caption{CMB temperature angular power spectrum for different $\xi_{\mathrm{g}} = G/G_N$.
Top left panel is obtained with a constant $\xi_{\mathrm{g}}$, whereas the top right panel is obtained with $\xi_{\mathrm{g}}\neq1$ for a redshift transition from $G_N$ at $z_{tr} > 8$.
Bottom left panel is obtained with $\xi_{\mathrm{g}}\neq1$ for $z_{tr} > 30$, whereas the bottom right panel is obtained with $\xi_{\mathrm{g}}\neq1$ with a transition width $\pm10$ around $z_{tr}\sim 30$.}
\label{fig:TTCell}
\end{figure}

\section{Datasets and analyse method}\label{sect:data_method}

\begin{figure}[!ht]
\centering
\includegraphics[scale=0.4]{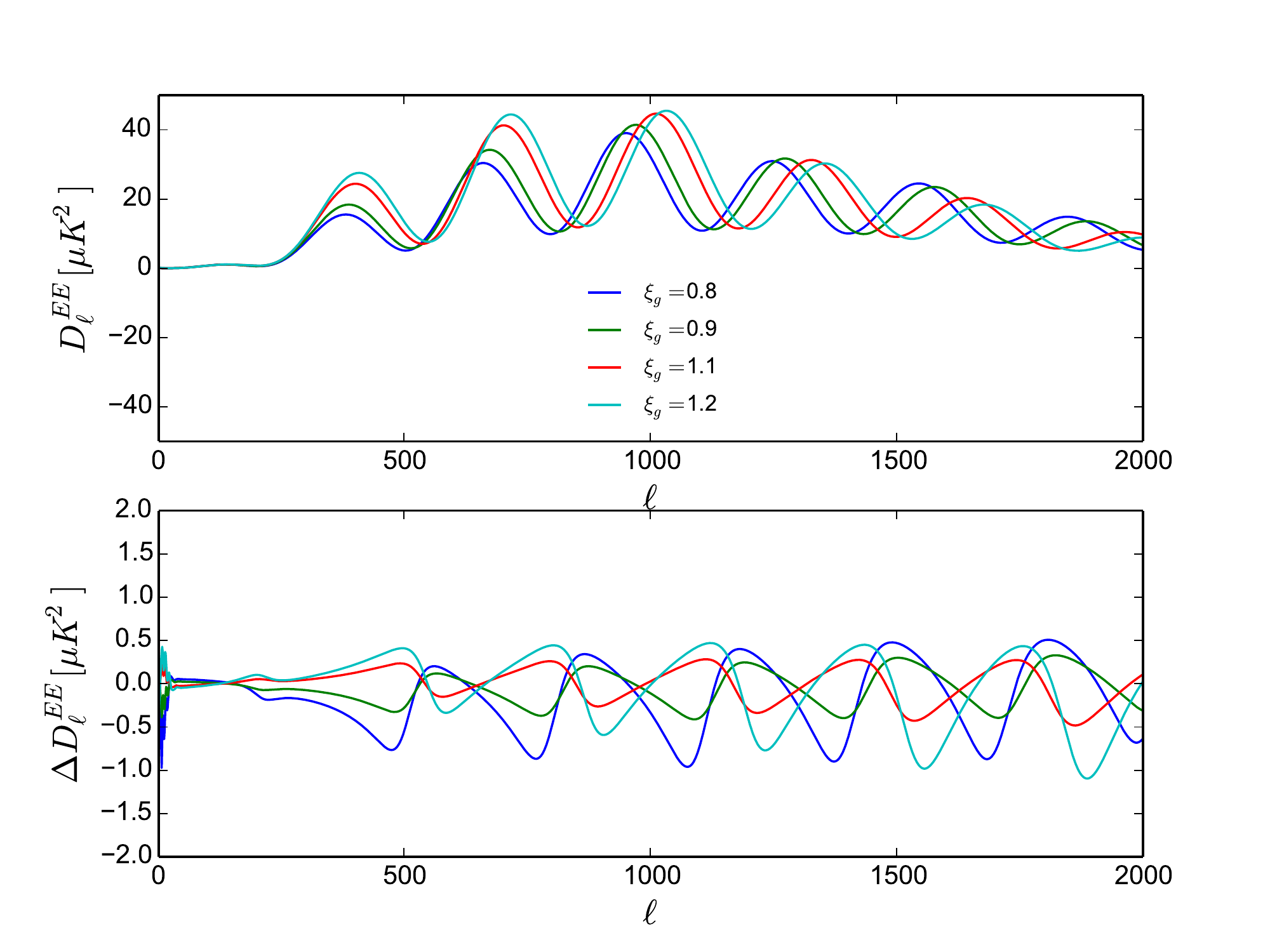}
\includegraphics[scale=0.4]{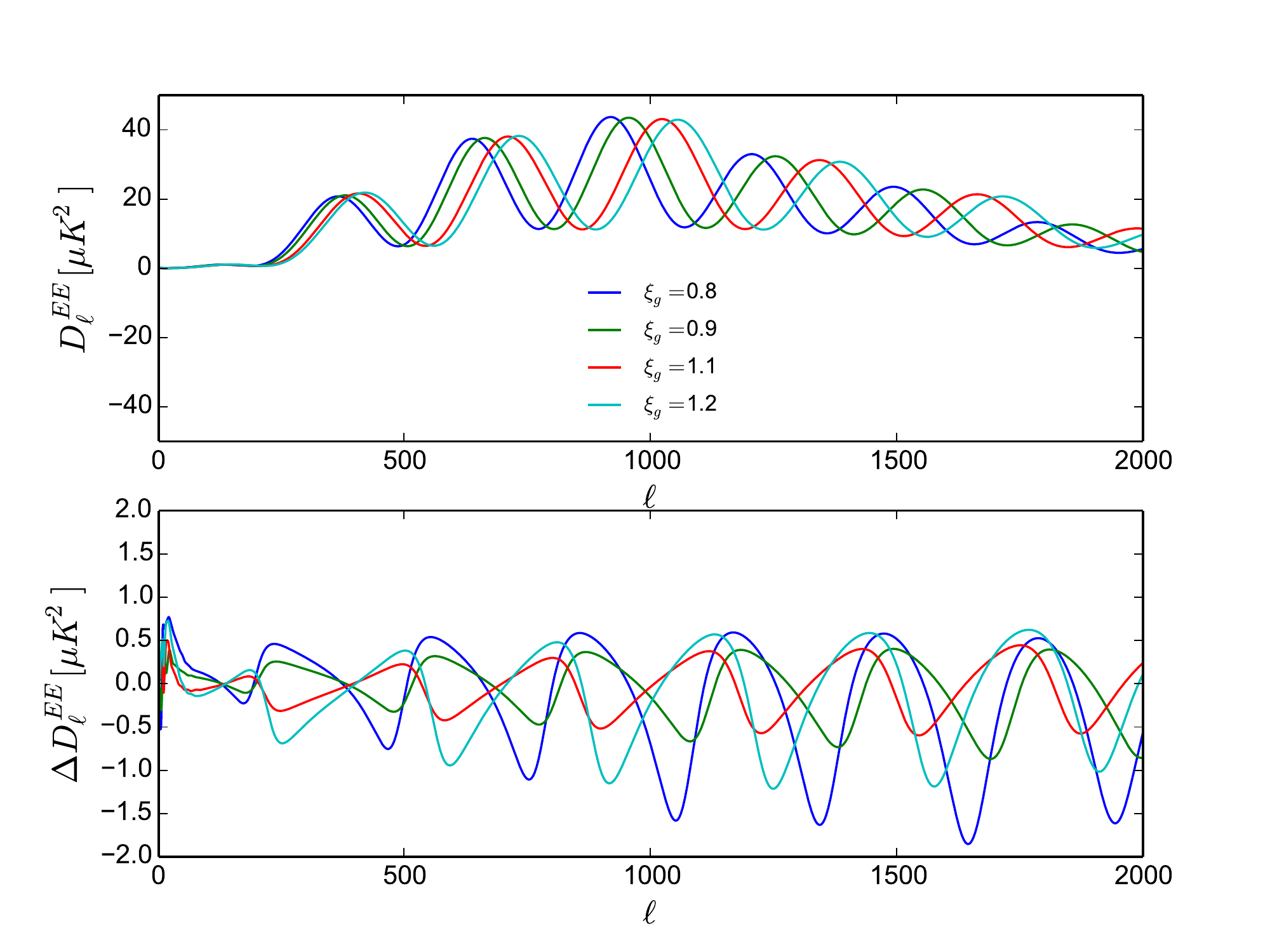}
\includegraphics[scale=0.4]{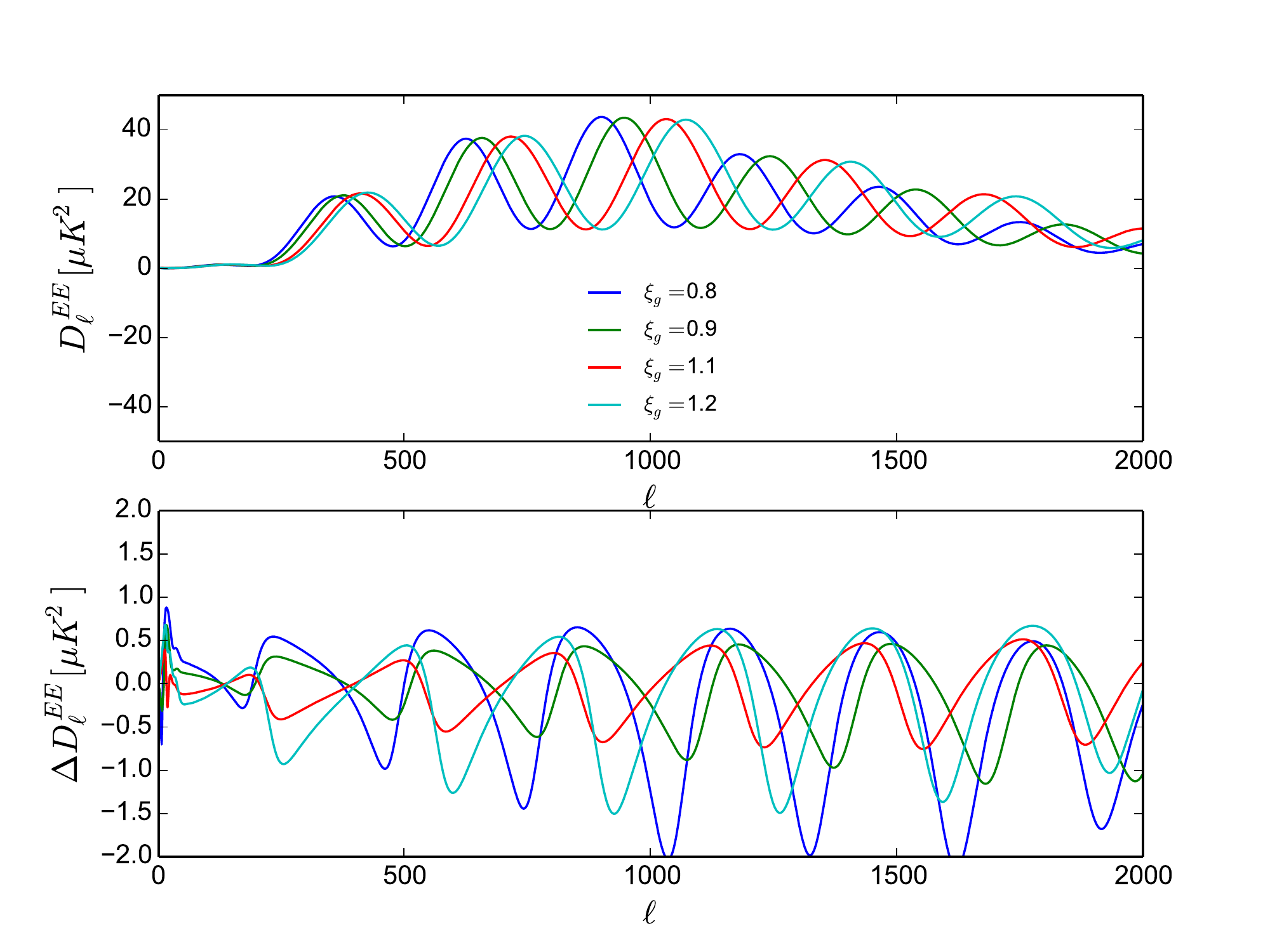}
\caption{CMB polarisation angular power spectrum for different $\xi_{\mathrm{g}} = G/G_N$ and their residuals with respect to fiducial $G_N$. Top: Obtained for the same change in $G$ at all redshifts. Middle: Obtained for a redshift transition from $G_N$ at $z_{tr} > 8$. Bottom: Obtained for a redshift transition from $G_N$ at $z_{tr} > 30$.} 
\label{fig:EECell}
\end{figure}

In order to implement our model, we modify the cosmological solver CLASS \cite{2011arXiv1104.2932L} that is able of calculating the temperature-temperature (TT) and polarisation-polarisation (EE) angular correlations and their cross correlations (TE) of the CMB, as well as the cosmological distances and the redshift epoch of the baryonic acoustic oscillation early occurrence, both needed to compare with galaxy clustering BAO measurements. 

We then perform a Monte Carlo Markov Chain (MCMC) analysis to constrain deviations of $G$ from fiducial CODATA value using MontePython \cite{Audren:2012wb}, a tool that embeds CLASS functions in the Planck likelihoods as well those needed for the BAO likelihood. For that, we use Planck $C^{\rm TT}_\ell$, $C^{\rm TE}_\ell$, and $C^{\rm EE}_\ell$ at high $\ell$ and TT and EE correlations at low $\ell$ from the final mission survey release of Planck (2018) \cite{Planck:2018vyg}, as well as the BAO measurements from Beutler et al. (2011) \cite{Beutler:2011hx}, Ross et al. (2015) \cite{Ross:2014qpa} and Alam et al. (2017) \cite{BOSS:2016wmc}.

Besides the free constant case, we choose the redshift threshold at which the model reverts back to constant fixed $G$ case, which allows us to distinguish between the different effects. We choose two different thresholds: the first one is $z_{tr} \sim 30$, which occurs later than the recombination epoch (around the beginning of formation of structures) but well before the reionisation epoch; the second one is around $z_{tr} \sim 8$, happening at the latter epoch but well before the BAO local times, since this last epoch would have been covered by the free constant case that goes close to our present times around $z\sim0.0025$. Even with the free constant case we do not consider redshifts below the latter value, to try evade solar and astrophysical constraints from non linear structure formation, since also $z\sim0.0025$ translates into the size of $\sim 10$ Mpc, close to that of our local galaxy cluster. In all these models the transition back from modified $G$ to the fiducial $G_N$ happens sharply. This could enhance the ISW effect. To test the impact of such effect, we consider the same binning but with a smooth transition with a large width following the model described by Eq.~(\ref{equ:ztrans}) in Sect.~\ref{sect:model_prob}.

To justify the above choices, we illustrate their effects on our observables. In Fig.~\ref{fig:TTCell}, we show the temperature correlations angular power spectrum $C^{\rm TT}_\ell$s for some of the different aferomentioned cases. On top left panel, we observe that the free constant case ($\xi_{\mathrm{g}} \neq 1$) affects the angular power spectrum at small scales. For the other cases, the effects are prominent at larger scales. We, thus, expect that the latter will put more constraints on deviations of $G$ from the fiducial. Furthermore, we observe that the dynamical $G$ affects strongly the ISW scales (see Sect.~\ref{sect:model_prob}), contrary to the case of constant $G$ where the variation has a monotonic increase. However, the influence of the dynamical $G$ gets smaller (in average) when the redshfit threshold gets close to late times (bottom left and bottom right panels). Thus, in order to take this effect into account, we compare the MCMC chains with and without the smooth transition.

In Fig.~\ref{fig:EECell}, we show the polarisation correlations angular power spectrum $C_\ell^{\rm EE}$s for the three cases: one with the change in $G_N$ for all redshifts, then, in the middle, with a transition threshold below $z_{tr}\sim 30$ and, at bottom, for $z_{tr}\sim8$. As described in Sec.~\ref{sect:model_prob}, we observe, especially in the last two panels, that the difference in amplitude of the $C_\ell^{\rm EE}$s is significant and it could reach twice its value even for a $10\%$ modification on value of $G$. This suggests that the probe could put strong constraints on the value of $G$.

\begin{figure}[!t]
\centering
\includegraphics[scale=0.5]{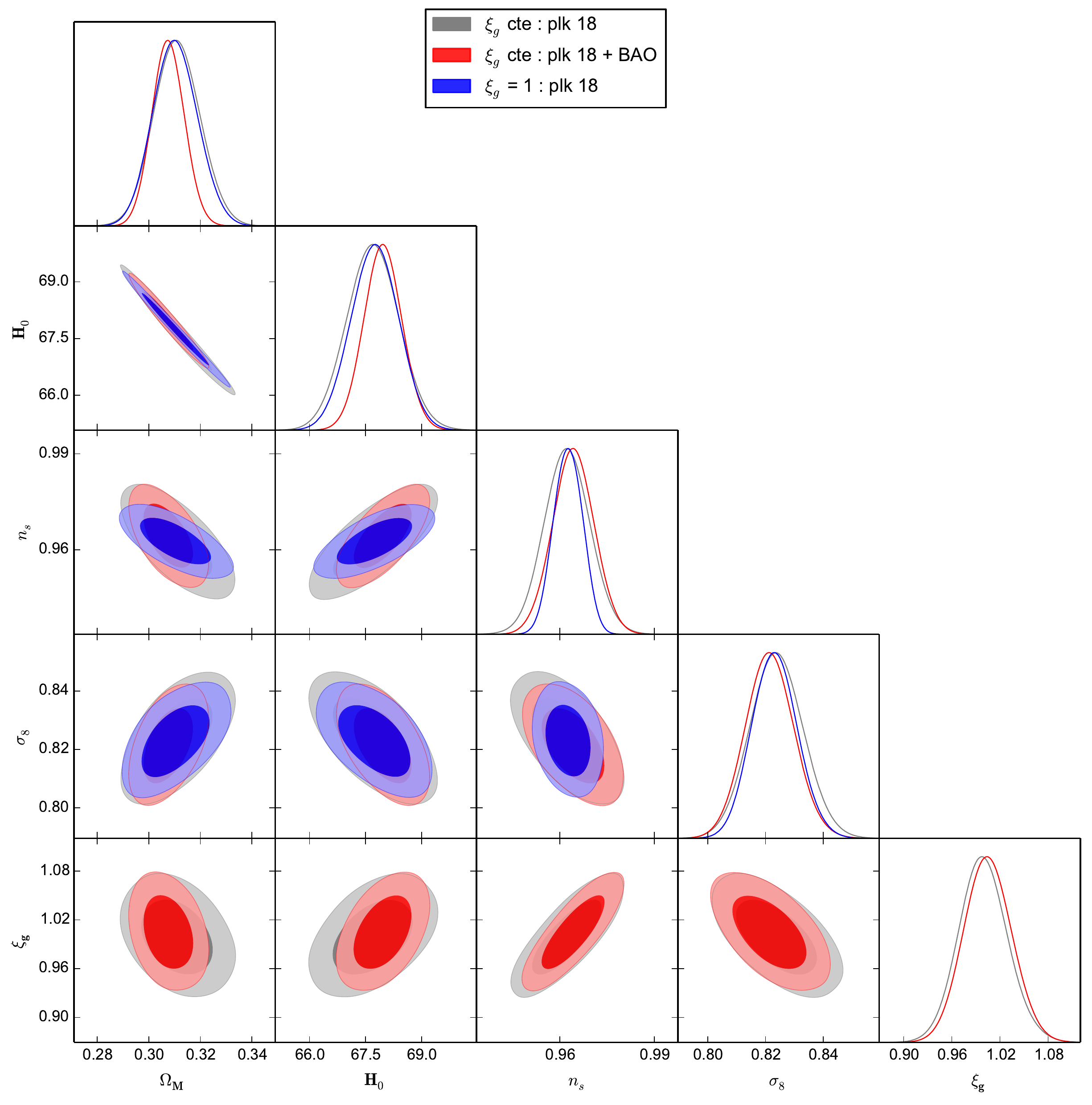}
\caption{Confidence contours (68 and 95\%) for cosmological parameters in the fixed CODATA $G_N$ case in comparison with a free constant $\xi_{\mathrm{g}} \neq 1$ using CMB data only or combined with BAO data.}
\label{fig:plkvsGctethermoreiofix}
\end{figure}

\section{Results and discussion}\label{sect:G_mod_results}

\begin{figure}[!t]
\centering
\includegraphics[scale=0.5]{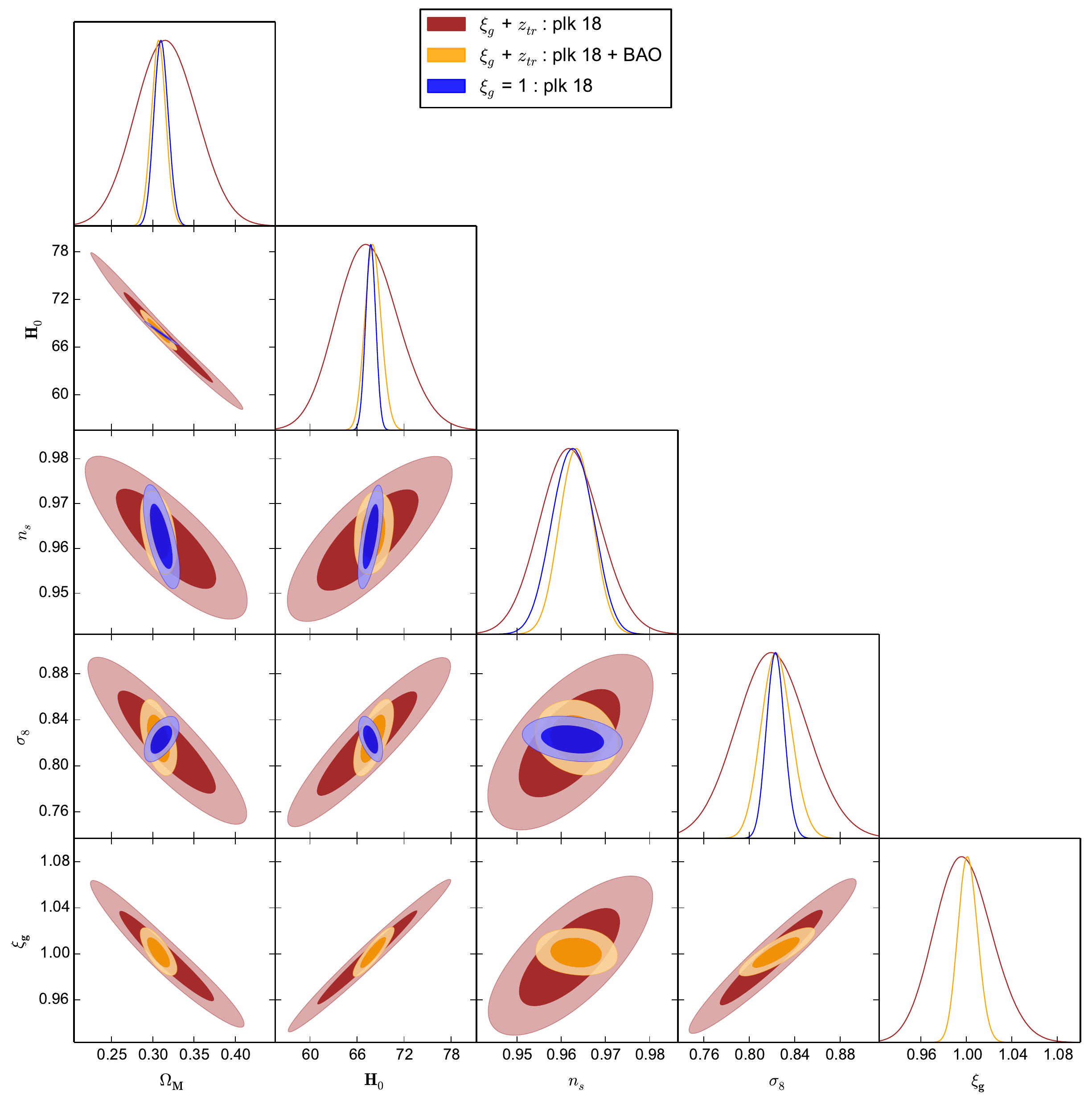}
\caption{Confidence contours (68 and 95\%) for cosmological parameters in the fixed CODATA $G_N$ case in comparison with a free constant $\xi_{\mathrm{g}}\neq1$ for a redshift transition from $G_N$ at $z_{tr} > 30$ using CMB data only or combined with BAO data.}
\label{fig:plkvsGscutthermoreiofix}
\end{figure}

In Fig.~\ref{fig:plkvsGctethermoreiofix}, we show the confidence contours of the cosmological parameters using the CMB temperature and polarisation correlations from Planck 2018 datasets with the free $G$ constant case, together with the case where $G$ is fixed to the CODATA value. We usually show the confidence values obtained at the 68\% level unless otherwise stated. We quantify the deviation with the parameter $\xi_{\mathrm{g}} = G/G_N $, where $\xi_{\mathrm{g}} = 1.0$ corresponds to $G$ taking the fiducial CODATA value. We first observe that the fiducial value of $G$ is within the inferred $\xi_{\mathrm{g}}$, with the maximum likelihood at $1.0\pm$0.04. This is the consequence of $\xi_{\mathrm{g}} \neq 1.0$ effect on the small scales of the CMB spectrum as illustrated in Sec.~\ref{sect:data_method} and in Fig.~\ref{fig:TTCell}, as a result of the correlation between $n_s$ and $\xi_{\mathrm{g}}$. However the discrepancy with $\sigma_8$ or $H_0$ remains high as the constraints do not change significantly with respect to the fixed $G_N$ case. We neither observe a change in $\xi_{\mathrm{g}}$ constraints when we combine CMB with BAO probes. This is due to the fact that a different constant fiducial value of $G$ only changes the cosmic clock without altering the BAO observable.

\begin{figure}[!t]
\centering
\includegraphics[scale=0.5]{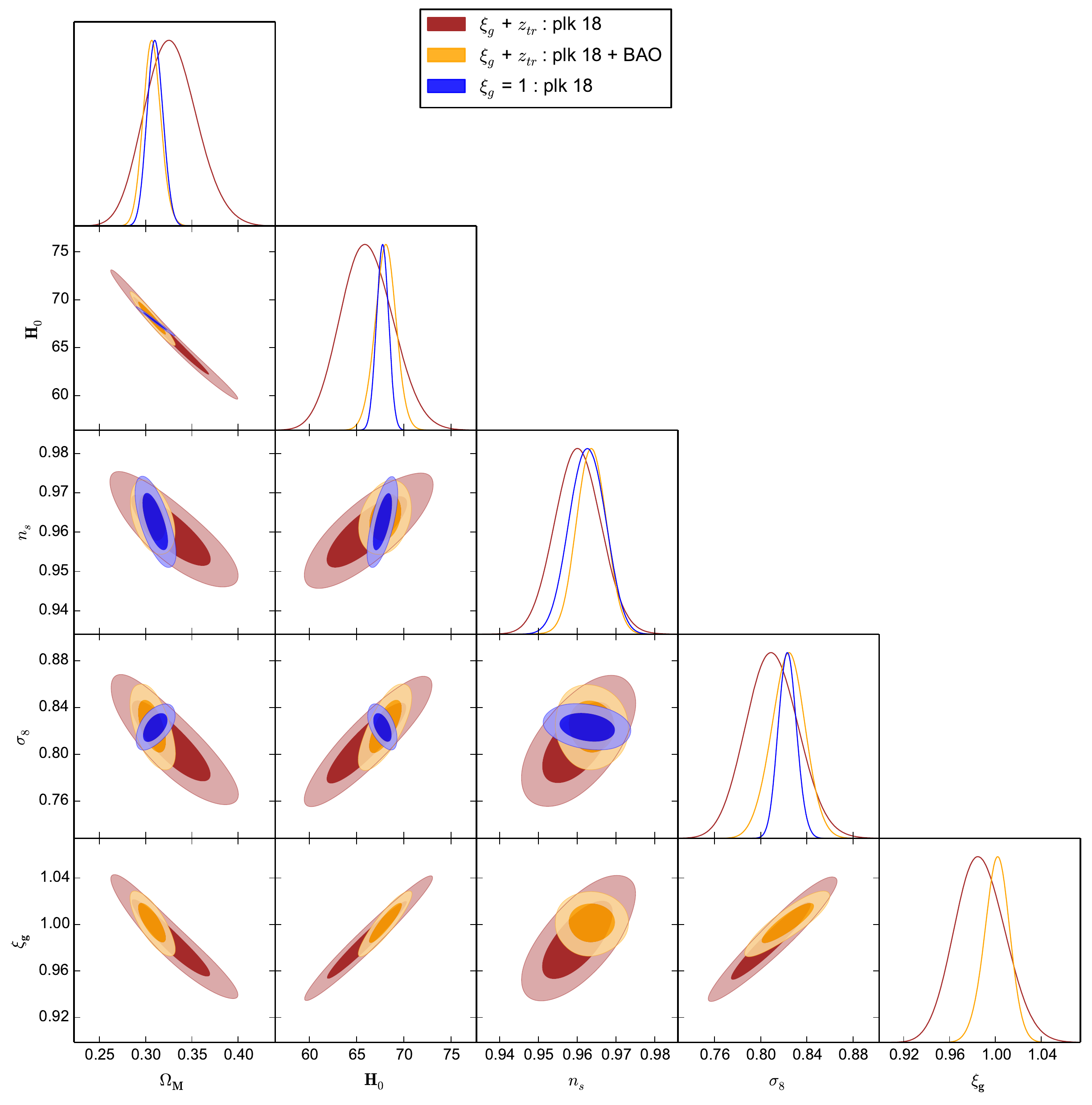}
\caption{Confidence contours (68 and 95\%) for cosmological parameters in the fixed CODATA $G_N$ case in comparison with a free constant $\xi_{\mathrm{g}}\neq1$ for a redshift transition from $G_N$ at $z_{tr} > 8$ using CMB data only or combined with BAO data.
}
\label{fig:plkvsGscutztrans=8thermoreiofix}
\end{figure}

In Fig.~\ref{fig:plkvsGscutthermoreiofix}, we show the confidence contours when we allow $\xi_{\mathrm{g}}$ to vary for $z_{tr}> 30$. We observe that, using Planck 2018 only, the constraints on $G$ are similar to those of the free constant $\xi_{\mathrm{g}}$ case, with the maximum likelihood at $\sim$ 1.0 and a 1$\sigma$ variation of $\pm$0.03. The reason is that, at lower $\ell$, the impact of $\xi_{\mathrm{g}}$ on TT correlation (Fig.~\ref{fig:TTCell} bottom left) is lower than the one observed with the free constant $\xi_{\mathrm{g}}$ on all redshifts case, while the opposite is seen for the effect on the polarisation correlations as shown in Fig.~\ref{fig:EECell}. For the Hubble parameter, we observe that the discrepancy is alleviated as the constraints become much broader than the constant $G$ case, with the local Hubble value of $\sim 73-74$ now falling within less than 1$\sigma$ of CMB inferred one. The $\sigma_8$ discrepancy is reduced in this parameterisation as well, and the value corresponding to that from local probes is reached at the 2$\sigma$ level. This is expected from the non monotonic effect of $\xi_{\mathrm{g}}$ on the ISW level seen in Sec.~\ref{sect:data_method} in Fig.~\ref{fig:TTCell}. However, the combination with the BAO datasets restores the discrepancy on both $\sigma_8$ and $H_0$. The BAO addition also tightens the constraints on $\xi_{\mathrm{g}}$ to $\pm0.03$ at 3$\sigma$, as a consequence of $\xi_{\mathrm{g}}$ that now needs to accommodate this paramerterization with two observations that disfavor a transition between the two redshifts.

\begin{figure}[!h]
\centering
\includegraphics[scale=0.5]{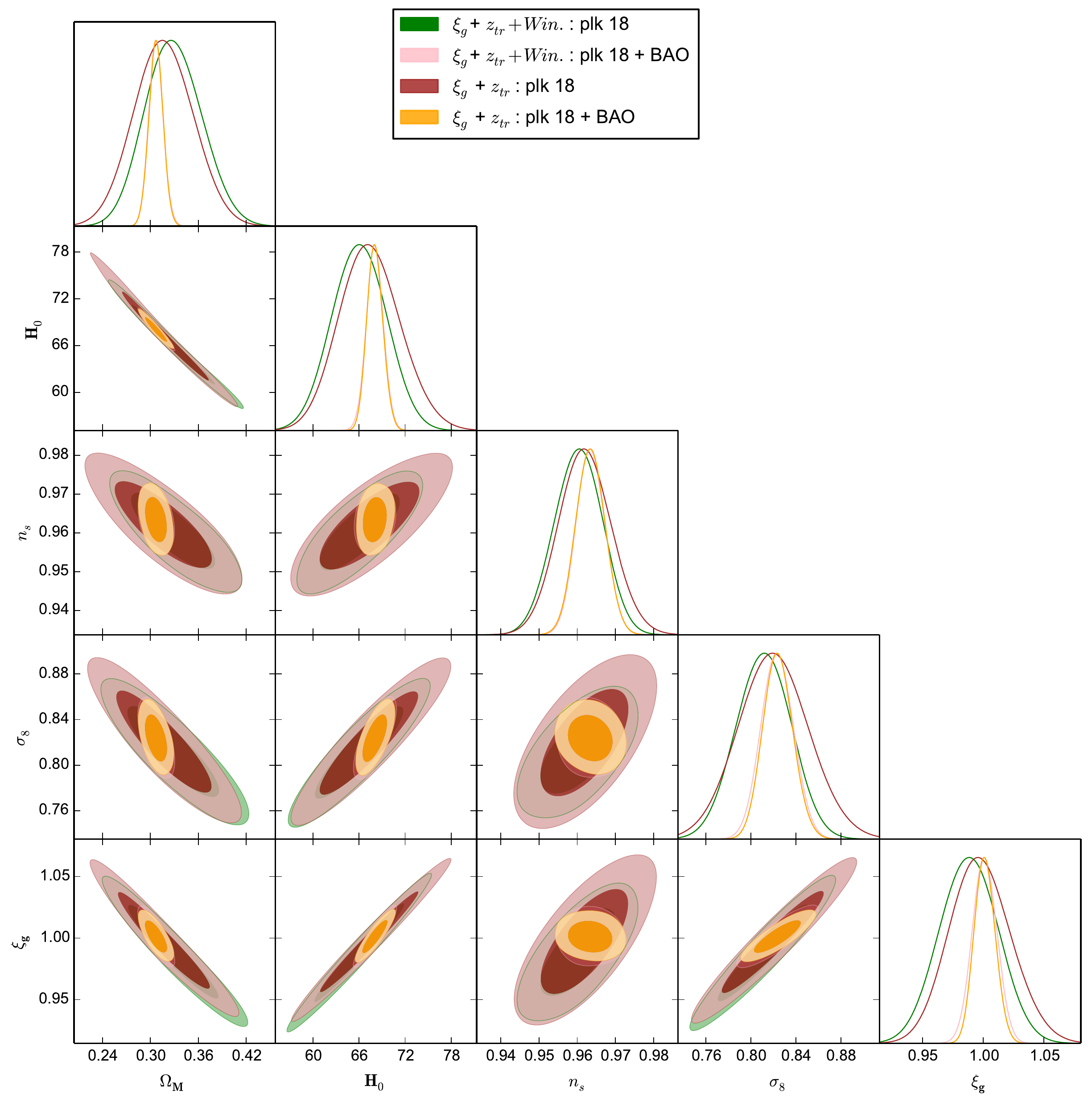}
\caption{Confidence contours (68 and 95\%) for cosmological parameters for a free constant $\xi_{\mathrm{g}}\neq1$ with a sharp transition from $G_N$ after $z_{tr} > 30$ in comparison with a transition width $\pm10$ around the same redshift threshold using CMB data only or combined with BAO data.}
\label{fig:4Gscutztrans=30thermoreiofixBAO}
\end{figure}

Assuming a value of $z_{tr}\sim 8$, we observe in Fig.~\ref{fig:plkvsGscutztrans=8thermoreiofix} that the constraints on $\xi_{\mathrm{g}}$ stay almost the same or get slightly tighter despite that $\xi_{\mathrm{g}}$ is allowed to deviate from its fiducial value for a larger range of redshifts. The reason here is that the sensitivity of the CMB polarisation spectrum to $\xi_{\mathrm{g}}$ is weaker than the $z_{tr} \sim 30$ case, as seen in Sect.~\ref{sect:data_method} in Fig.~\ref{fig:EECell}. The same discussion applies to $\sigma_8$ or $H_0$ constraints with the only difference that now BAO data are unable to completely restore the discrepancy as in the previous case because $G$ is allowed to vary for a much longer period. 
\begin{figure}[!t]
\centering
\includegraphics[scale=0.5]{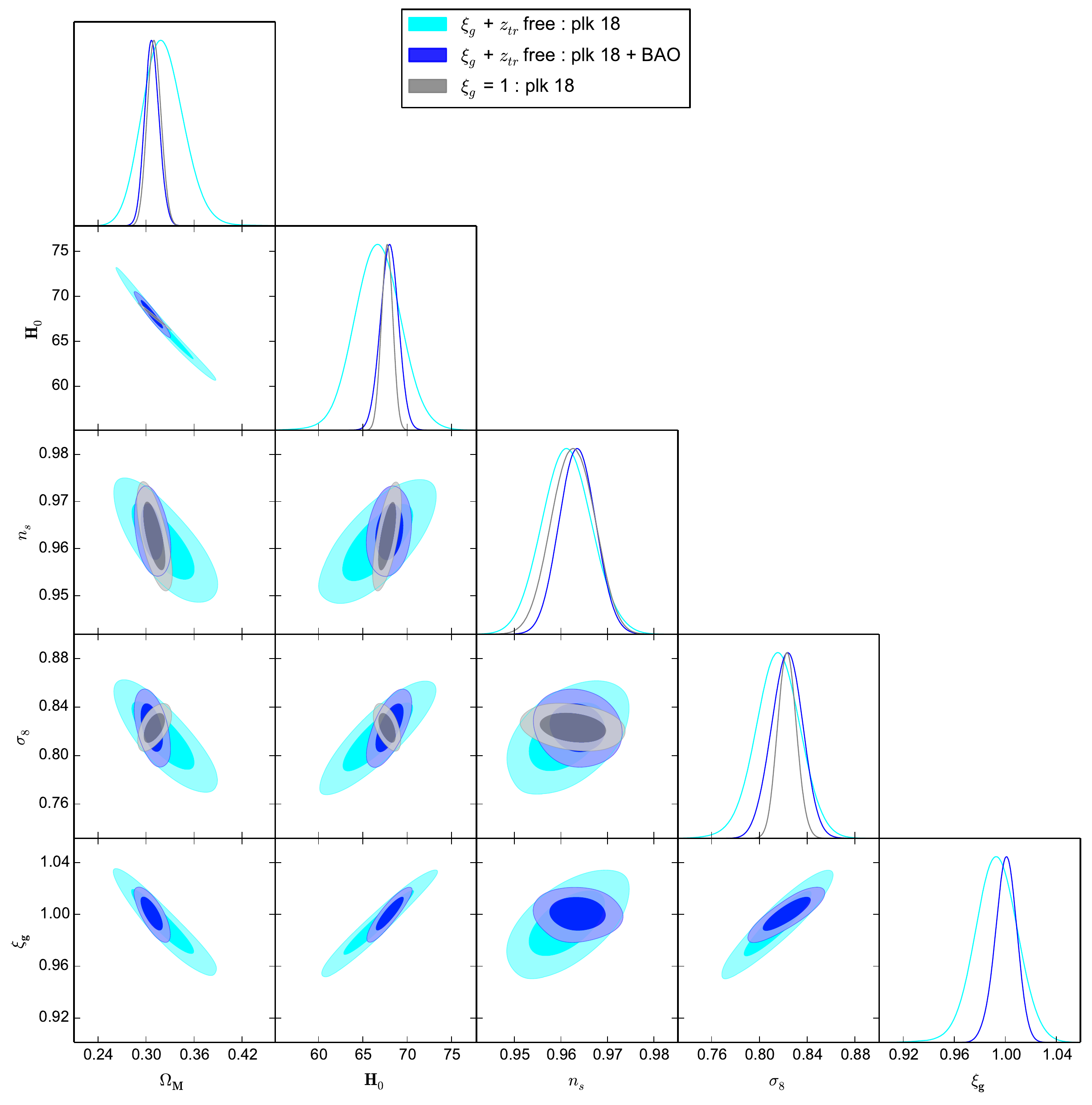}
\caption{Confidence contours (68 and 95\%) for cosmological parameters in the fixed CODATA $G_N$ case in comparison with a free constant $\xi_{\mathrm{g}}\neq1$ and a free redshift transition threshold from $G_N$ using CMB data only or combined with BAO data.}
\label{fig:plkGthermoreiofixBAOzrange}
\end{figure}

\begin{figure}[!t]
\centering
\includegraphics[scale=0.5]{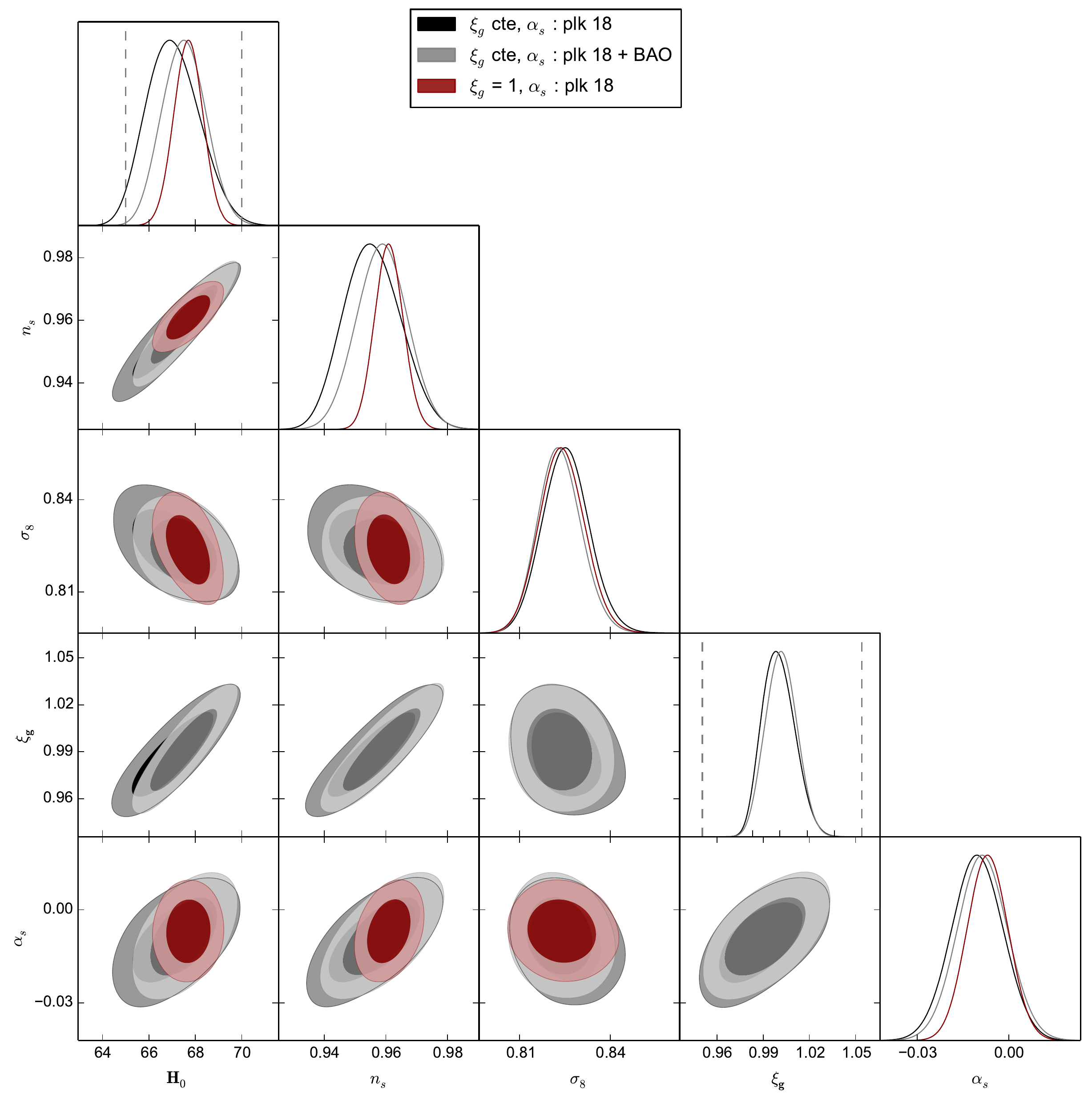}
\caption{Confidence contours (68 and 95\%) using CMB data only or combined with BAO data for cosmological parameters in the case of a free $\alpha_s$ and a fixed CODATA $G_N$ in comparison with a free $\alpha_s$ along with $\xi_{\mathrm{g}}\neq 1$.}
\label{fig:plkvsGztransthermoreioalphaBAO}
\end{figure}

\begin{figure}[!t]
\centering
\includegraphics[scale=0.5]{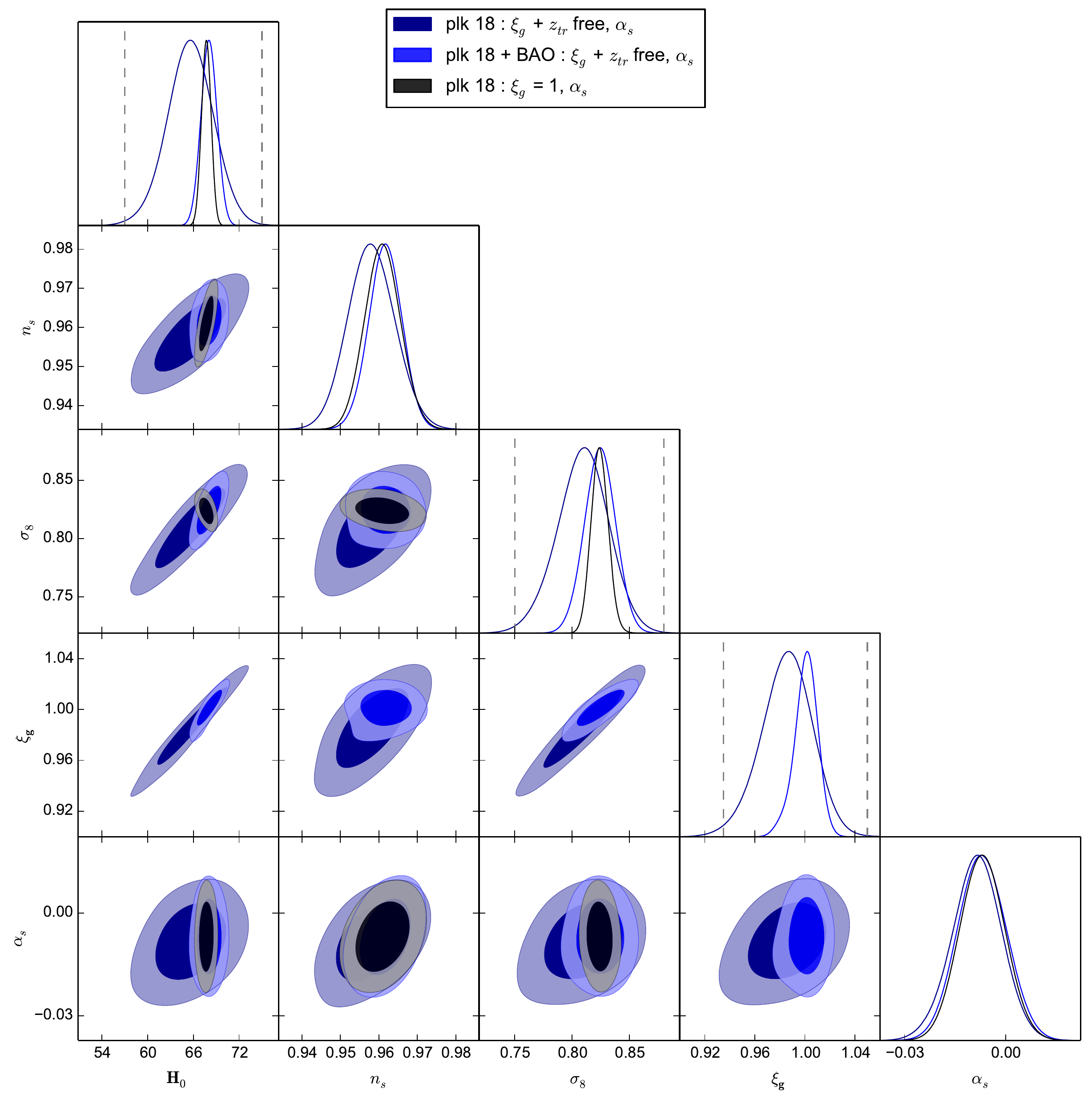}
\caption{Confidence contours (68 and 95\%) using CMB data only or combined with BAO data for cosmological parameters for a free $\xi_{\mathrm{g}}$ with a free redshift threshold and a free $\alpha_s$ together with a free $\alpha_s$ but with a fixed CODATA $G_N$.}
\label{fig:plkvsGztransthermoreioalphaBAO-bs}
\end{figure}

So far we have considered models that deviate sharply at some chosen redshift; however it is interesting to consider smooth transitions that mimics a linear variation around the threshold redshift. In Fig.~\ref{fig:4Gscutztrans=30thermoreiofixBAO} we show the results combining CMB and BAO data when using a $\tanh$ function which allows for the transition at $z_{tr} \sim 30$. We observe no substantial impact on all the parameters when we combine CMB to BAO  data, indicating that the ISW effect is not enough to constrain deviations of $G$. We also verified (without showing here the results) the case where $z_{tr} \sim 8$ and we found an even smaller difference between the confidence contours of the sharp transition parameterisation and the smoothed one.

Finally, since we found that the constraints could change with the redshift transition threshold, we decided, in a more model independent approach, to repeat the bayesian analysis with $z_{tr}$ as an additional free parameter. Using Planck 2018 data alone, we observe in Fig.~\ref{fig:plkGthermoreiofixBAOzrange} that the $\xi_{\mathrm{g}}$ maximum likelihood coincides with the fiducial $G_N$ one with much tighter constraints then the other cases to $\pm0.02$. The $H_0$ and $\sigma_8$ tensions are alleviated but after adding BAO data, the discrepancy on $H_0$ is restored and the ability to fix the $\sigma_8$ discrepancy is reduced. The addition of BAO also tighten the constraints on $\xi_{\mathrm{g}}$ with 68$\%$ confidences values at $\pm0.01$.

Besides the redshift threshold influence, we noticed that the spectral index $n_s$ shows a correlation with $\xi_{\mathrm{g}}$ in all the cases considered in this work. Since the CMB probe puts strong constraints on $n_s$, we decided to relax the latter by allowing a scale dependency through the spectral index running parameter $\alpha_s = \rmd \rm{n}_s / \rmd \rm{ln} k$. In Figs.~\ref{fig:plkvsGztransthermoreioalphaBAO} and~\ref{fig:plkvsGztransthermoreioalphaBAO-bs} we show the confidence contours on the cosmological parameters including a free running spectral index different from the null fiducial value in the fixed CODATA $G_N$ case along with two of the parmetrisations for $\xi_{\mathrm{g}}$ with opposite behavior in terms of their redshift transition value: the first considering a free constant $\xi_{\mathrm{g}}$ different from fiducial at all redhsifts (i.e. with a threshold fixed to $z_{tr} \sim 0.0 $), while the second allows a free $z_{tr}$ on a very large range spanning from nowadays until the recombination epoch. For the case of a constant free $G$ and a free $\alpha_s$ (Fig.~\ref{fig:plkvsGztransthermoreioalphaBAO}), we observe that the running is correlated with the $\xi_{\mathrm{g}}$ with the later confidence contours tighter than the case with vanishing $\alpha_s$ (represented by the dashed vertical lines). This comes from the fact that the data rather prefers a negative $\alpha_s$ instead of larger deviations of $\xi_g$. The latter constraints are however slightly broader than the $\xi_g =1$ case, resulting in a larger constraints on the Hubble parameter with respect to the previous case with a vanishing $\alpha_s$, represented by the dashed vertical lines in the $H$ single likelihood box. We finally note that the confidence contours remain almost the same for all the parameters when combining with BAO. In the case of allowing the redshift transition threshold also to vary (Fig.~\ref{fig:plkvsGztransthermoreioalphaBAO-bs}), we observe that the introduction of the running parameter does not have much impact on the different constraints, since already in the previous case relaxing $z_{tr}$ has showed a preference for a tighter $\xi_g$ though the deviation of the $\alpha_s$ from the null fiducial value translates into slightly wider constraints on the $\xi_g$, $\sigma_8$ or $H_0$ (with the previous represented by the vertical dashed lines in each box), with the last two parameters again restored to their fiducial values when we combine with BAO.

\section{Conclusions}\label{sect:G_mod_concl}

In this paper we have studied the impact of a varying $G$ onto some of the cosmological observables at both macro or micro physical scales. For that, we used CMB temperature and polarisation correlations data alone or in combination with BAO distance measurements from galaxy clustering. We performed a MCMC analysis allowing a constant free $G$ different from fiducial or a dynamical $G$ that switches value at a given redshift. The objective of this analysis was to update the constraints on the deviation of $G$ from $G_N$ and investigate whether the discrepancy on the matter fluctuation $\sigma_8$ and $H_0$ parameters could be alleviated by these parameterisations. 

For a free constant $G$ case, we found that this parameterisation has no effect on fixing the $\sigma_8$ or $H_0$ discrepancy while the $\xi_{\mathrm{g}} = G / G_N$ is constrained to 1.0$\pm0.04$. Then, to separate constraints coming from lower redshifts from the ones at recombination epoch, we considered a dynamical $G$, with two redshift threshold at which $G$ transits from $G_N$, one at high redshift $z_{tr} \sim 30 $ and another at $z_{tr} \sim 8 $. With both redshift transitions, we found that we were able of alleviating the tensions on $H_0$ and $\sigma_8$ only if we use the CMB temperature and polarisation correlations alone and at the expense of wider constraints, while the discrepancies were restored when we combine with BAO in the higher redshift transition option or only reduced with the lower redshift transition case. We also found that the constraints remain almost unchanged whether we consider a sharp or a smooth wide transition of the $G$ from its fiducial value at the given redshift, as an indication that the ISW effect is not enough constrained by CMB data due to the large cosmic variance present at large scales in the CMB correlations measured.

Finally, we considered two additional more general cases: one allowing a free redshift threshold in a further model independent approach, and the other introducing another degree of freedom by a non vanishing running of the spectral index, since the later was found in correlation with $\xi_{\mathrm{g}}$. In the first case, we found that we are still not able of alleviating the tension on the Hubble constant when the BAO are combined with CMB data, while the discrepancy on $\sigma_8$ is reduced and $\xi_{\mathrm{g}}$ is constrained to 1.0$\pm0.01$. In the case where we allowed a free running spectral index $\alpha_s$, we found that the constraints on $\xi_{\mathrm{g}}$ are tightened with the data rather preferring a negative $\alpha_s$, with only a small improvement of alleviating the discrepancies on $H_0$ and $\sigma_8$ was found. When we considered a free redshift transition value along with a free running of the spectral index, we observed that the data is now preferring a stronger correlation $\xi_{\mathrm{g}} - \alpha$ than $\xi_{\mathrm{g}} - n_s$ allowing further deviation of $\xi_{\mathrm{g}}$ from the fiducial and an alleviation of the tensions but only if BAO observations are not included. We conclude that there remains a preference to the fixed CODATA $G_N$ model and that the different parameterisations we considered are not able of alleviating the Hubble and the matter fluctuation parameter discrepancies.

\section*{Acknowledgements}
DS acknowledges financial support from the Fondecyt Regular project number 1200171. ZS acknowledge fruitful discussions with Brahim Lamine.

\bibliographystyle{JHEP}
\bibliography{G_var}

\end{document}